\documentclass[11pt,a4paper]{article}
\usepackage{jheppub}

\usepackage{amsmath,amssymb}
\usepackage{graphicx}
\usepackage{algorithmic,algorithm}
\usepackage{breakurl}
\usepackage{color}

\newcommand{\reduze}{{\tt Reduze}}
\newcommand{\fermat}{{\tt Fermat}}
\newcommand{\ginac}{{\tt GiNaC}}
\newcommand{\qgraf}{{\tt QGRAF}}

\newcommand{\berkeleydb}{{\tt Berkeley\ DB}}
\newcommand{\mathematica}{{\tt Mathematica}}
\newcommand{\maple}{{\tt Maple}}
\newcommand{\form}{{\tt FORM}}
\newcommand{\yaml}{{\tt YAML}}
\newcommand{\diana}{{\tt DIANA}}
\newcommand{\feynarts}{{\tt FeynArts}}

\definecolor{shadecolor}{rgb}{.95,.95,1}
\definecolor{framecolor}{rgb}{.1,.0,.7}
\definecolor{emphcolor}{rgb}{.1,.0,.7}

\makeatletter
\renewcommand\@fpheader{\hfill \parbox{3cm}{ZU-TH 01/12\\ MZ-TH/12-03 \\ BI-TP 2012/02}}
\renewcommand\@journal{}
\makeatother

\title{Reduze 2 -- Distributed Feynman Integral Reduction}
\collaborationImg{\includegraphics[scale=0.75]{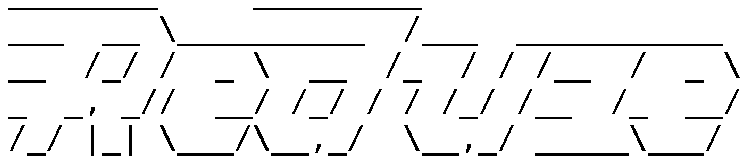}}
\author[a,b]{A.~von~Manteuffel,}
\author[c]{C.~Studerus}

\affiliation[a]{
  Institut f\"ur Theoretische Physik,
  Universit\"at Z\"urich,
  Winterthurerstrasse 190,\\
  CH-8057 Z\"urich, Switzerland}
\affiliation[b]{
  Institut f\"ur Physik (THEP),
  Johannes Gutenberg-Universit\"at,
  D-55099 Mainz, Germany}
\affiliation[c]{
  Fakult\"at f\"ur Physik,
  Universit\"at Bielefeld,
  Postfach 100131,
  D-33501 Bielefeld, Germany}

\emailAdd{manteuffel@uni-mainz.de}
\emailAdd{cedric@physik.uni-bielefeld.de}

\abstract{
\reduze\ is a computer program for reducing Feynman integrals to
master integrals employing a variant of Laporta's reduction algorithm.
This article describes version 2 of the program.
New features include the distributed reduction of single topologies
on multiple processor cores.
The parallel reduction of different topologies is supported
via a modular, load balancing job system.
Fast graph and matroid based algorithms allow for the identification of
equivalent topologies and integrals.
}

\begin{document}

\maketitle

\hyphenation{Stu-de-rus}


\section{Introduction}

In perturbative quantum field theory, the traditional method to compute
cross sections and distributions for a physical process involves generating
tree and loop amplitudes via Feynman diagrams and interfering them.
Simplifications of the expressions are performed at the analytical level.
Here an essential part is the reduction of the typically dimensionally regularized
loop integrals~\cite{'tHooft:1972fi} to a small number of standard integrals.
This step can be performed at the amplitude level for tensor integrals or,
after contraction of Lorentz indices, at the level of interferences for scalar
integrals. Considering the case of scalar integrals,  
integration by parts (IBP) identities~\cite{Tkachov:1981wb,Chetyrkin:1981qh}
and Lorentz invariance (LI) identities~\cite{Gehrmann:1999as} may be used
for a systematic reduction to a set of independent integrals, called master
integrals.
The standard reduction algorithm by Laporta~\cite{laporta} defines an
ordering for Feynman integrals, generates identities and solves the
resulting system of linear equations.
Alternative methods to exploit IBP and LI identities for reductions have been
proposed~\cite{Smirnov:2005ky,Smirnov:2006tz,Gluza:2010ws,Schabinger:2011dz},
see also \cite{Lee:2008tj,Grozin:2011mt} and references therein.
Public implementations of different reduction algorithms are available with the
computer programs AIR~\cite{Anastasiou:2004vj}, FIRE~\cite{Smirnov:2008iw}
and the first version of \reduze~\cite{Studerus:2009ye}.

This article presents the new public reduction program \reduze\;2.
It is written in {\tt C++} and represents a major rewrite and extension of its
predecessor \reduze. In the following, the name \reduze\ refers to the new
version presented here.

In \reduze, integrals are indexed by integral families (``auxiliary topologies'')
and sectors (``topologies'') therein.
For the reduction, the program implements a fully distributed variant of
Laporta's algorithm using the Message Passing Interface (MPI).
In this way, not only different sectors can be reduced in parallel, but also
the integrals of a single sector can be reduced in a distributed computation.

The program allows to utilize multiple integral families within a calculation.
Special emphasis has been placed on finding relations between sectors
of the same or different integral families and employing them
to eliminate integrals.
Besides a straightforward combinatorial matcher, the program implements
graph and matroid theory based algorithms to compute such relations,
taking into account possible crossings of external momenta.
Similar to the program \diana~\cite{Tentyukov:1999yq}, \reduze\ may be used
to shift loop momenta of Feynman diagrams generated by a program like
\qgraf~\cite{qgraf} or \feynarts~\cite{Hahn:2000kx} to match sectors of
integral families.

Other features include the generation of differential equations for Feynman
integrals and the computation of (bare) amplitude interferences up to master
integrals, starting from Feynman diagrams generated by \qgraf.
For storing intermediate results of a reduction, optionally,
the transitional open source database \berkeleydb~\cite{berkeley}
can be used.

For the normalization of algebraic coefficients
in the identities, one can choose between \ginac~\cite{Bauer:2000cp}
and \fermat~\cite{fermat}.
Reduction identities and other results can be exported to
\form~\cite{form}, \mathematica~\cite{mathematica}, and \maple~\cite{maple}
format.
Configuration and job files use the \yaml\ format~\cite{yaml} and are parsed
with the {\tt yaml-cpp} parser~\cite{yamlcpp}.

\reduze\;2 was used to calculate the two-loop leading color corrections
to heavy-quark pair production in the gluon fusion channel~\cite{Bonciani:2010mn}.
Last but not least, \reduze\ is published as open source under the
{\tt GNU} General Public License (GPL) v3 and has no mandatory dependencies
on proprietary software.

\section{Basic concepts and notations}

\subsection{Integral families, sectors, and integrals}
\label{sec:fam_sec_int}

A \emph{propagator} $P$ is defined as the expression $1/(q^2 - a)$
where $q$ is a four-momentum and $a$ is constant.
The momentum $q$ of a propagator (defined up to a minus sign) is a
linear combination of loop momenta $k_i$ and external momenta $p_i$,
and $q^2$ is a scalar product in Minkowski space with the metric convention
$g = \mbox{diag}(1, -1, -1, -1)$.
In \reduze, also generalized propagators $1/(q l - m^2)$
with the scalar product of two different momenta $q$ and $l$ are available
to support more general irreducible numerators.

An $l$-loop \emph{integral family} (or ``auxiliary topology'') $F$ is an
ordered set $\{ P_1, \ldots, P_n \}$
of propagators $P_i$, $i=1,\ldots,n$, which is minimal and complete in the sense,
that any scalar product of a loop momentum $k_i$ with a loop momentum $k_j$
or an external momentum $p_j$ can be uniquely expressed as a linear combination
of inverse propagators and kinematic invariants.
Denoting the number of independent
external momenta by $m$, an integral family must contain exactly
$l\,(l+1)/2+l\,m$ propagators, where the first term counts the scalar products
between loop momenta only and the second term the products involving both loop and
external momenta.
A new feature of \reduze\  is its ability to handle several integral families
simultaneously.

A selection of $t$ propagators of an integral family defines a \emph{sector} of
this family.
Assuming a sector has the propagators $P_{j_1},\ldots, P_{j_t}$ with
$\{j_1,\ldots,j_t\} \subset \{1,\ldots,n\}$, then its identification number
is defined as
\begin{equation}
 ID = \sum_{k=1}^{t} 2^{j_k-1} \, .
\end{equation}
There are in general $\binom{n}{t}$ different $t$-propagator sectors and
$\sum_{t=0}^{n} \binom{n}{t} = 2^n$ sectors are contained in an integral
family. Their identification numbers fulfill $0 \leq ID \leq 2^n - 1$ .

A sector whose propagators form a subset of the propagators of another sector
of the same integral family is a \emph{subsector} of the other sector.

The purpose of an integral family is to index scalar loop integrals.
To every $t$-propagator sector with propagators $P_{j_1}$, $\ldots$, $P_{j_t}$
belongs a infinite set of $d$-dimensionally regularized $l$-loop
integrals~\cite{'tHooft:1972fi} which all share the same propagators.
These integrals have the generic form
\begin{equation}\label{eq:dim_reg_int}
I = \int \mbox{d}^d k_1 \ldots \int \mbox{d}^d k_l \,
P_{j_1}^{r_1} \ldots P_{j_t}^{r_t}
     P_{j_{t+1}}^{-s_1} \ldots P_{j_n}^{-s_{n-t}}
\end{equation}
with integer exponents $r_i \geq 1$ and $s_i \geq 0$.
In \reduze\  such an integral is represented by
\begin{equation}\label{eq:int_reduze_format}
\mbox{I}(F, t, ID, r, s,\{v_1,\ldots,v_{n}\})
\end{equation}
where $F$ denotes the integral family,
$r=\sum_{i=1}^t r_i \geq t$, $s=\sum_{i=1}^{n-t} s_i \geq 0$ and
$v_i$ is the exponent of propagator $P_i$.
Positive $v_i$ denote powers of regular propagators (non-trivial denominator),
negative $v_i$ denote powers of inverse propagators (non-trivial numerator),
and zero means absence of a propagator.
The numbers $t$, $r$, $s$ as well as the identification number $ID$ of the
sector, to which the integral belongs, can be calculated from the vector $v$.

Consider a $t$-propagator sector of a $n$-propagator integral family.
The number of integrals that one can build for certain values of $r$ and $s$ is given by
$
{\mathcal N}(n,t,r,s) = \binom{r-1}{t-1} \binom{s+n-t-1}{n-t-1} \, .
$
The two binomial factors count all possible ways to arrange the exponents of
the propagators in the denominator and numerator, respectively.

The integral with $r=t$ and $s=0$ of some sector is called 
\emph{corner integral} of this sector.

\subsection{Integration by parts (IBP) identities}
In dimensional regularization~\cite{'tHooft:1972fi} the integral over a total derivative
is zero. Let ${\mathbf I'}$ be the integrand of an integral of the form (\ref{eq:dim_reg_int}).
Then, working out the differentiation in
\begin{equation}
 \int \mbox{d}^d k_{i} \, \frac{\partial}{\partial k_{i}^{\mu}} \, \big[ q^{\mu} \,
 {\mathbf I'}(p_1, \ldots, p_{m}, k_1, \ldots, k_l) \big] = 0
\end{equation}
leads to the integration by parts (IBP) identities~\cite{Tkachov:1981wb,Chetyrkin:1981qh}.
The momentum $q$ is an arbitrary loop or external momentum. The index $\mu$ is summed over
but the index $i$ is not. If there are $l$ loop momenta and $m$ independent external
momenta one can therefore build $l\, (l+m)$ equations from one integral, the
\emph{seed integral}.

\subsection{Lorentz invariance (LI) identities}
One can also use the Lorentz invariance of the integrals~\cite{Gehrmann:1999as}. Taking
an integral ${\mathbf I}(p_{1},\ldots ,p_{m})$ the following equation holds
\begin{equation}
\sum_{n=1}^m \Big( p_{n}^{\nu} \frac{\partial}{\partial p_{n \mu}}
    - p_{n}^{\mu} \frac{\partial}{\partial p_{n \nu}} \Big)\, {\mathbf I}
    (p_{1},\ldots ,p_{m}) = 0 \, .
\end{equation}
The derivatives can be shifted directly to the integrand of the integral ${\mathbf I}$.
This equation can be contracted with all possible antisymmetric combinations of the
external momenta, e.g.\ $p_{1 \mu} p_{2 \nu} - p_{1 \nu} p_{2 \mu}$, which leads to
$m\,(m-1)/2$ equations where $m$ denotes the number of independent external momenta.
As it was shown in~\cite{Lee:2008tj} the LIs do not give new linearly independent
equations in addition to the IBPs. However, they can accelerate the convergence in a
reduction, since in general an LI identity generated from one seed integral cannot be
reproduced with the IBP identities generated from the same seed integral alone.
\reduze\  offers the possibility to use the LIs.

\subsection{Zero sectors}

It is possible that a whole sector is zero which means that all integrals belonging
to this sector are zero. A sector of an $l$-loop integral family is trivially
zero if it does not allow for a selection of $l$ propagator momenta which are
independent with respect to the $l$ loop momenta.
The graph based methods in \reduze, see section \ref{sec:physical_secs},
will automatically detect these cases.
As a second method, a sector is set to zero if the reduced IBP identities
generated from the seed integrals of this sector with $r=t$ and
$s=0, 1$ show that its corner integral is zero.

\subsection{Sector relations} \label{sec:sec_rel}

Given a scalar loop integral as well as one or several integral families,
suppose it is possible to map the integral to a linear combination of indexed
integrals of type \eqref{eq:int_reduze_format}.
In general, such a map is not unique.
Ambiguities may arise if sectors from different integral families have the same
set of propagators or if a transformation of loop and external momenta in
\eqref{eq:dim_reg_int} leads to a different linear combination of type
\eqref{eq:int_reduze_format}.
For the corner integral of some sector $S$ written in the form
\eqref{eq:dim_reg_int},
consider the transformation of integration variables
\begin{align}\label{eq:shift_trafo}
k_i &\to \sum_{j=1}^{l} M_{ij} k_j + \sum_{j=1}^{m} N_{ij} p_j
\end{align}
with $|\det{M}|=1$.
If the new integrand factors can be identified with propagators of a
sector $S'$, the ``shift'' transformation \eqref{eq:shift_trafo} defines a
\emph{sector relation} between $S$ and $S'$.
In this case, any integral in the sector $S$ can be expressed as a linear
combination of integrals in the sector $S'$ and subsectors of $S'$.
If $S$ and $S'$ are different, such a relation can be used to eliminate
one of the two sectors completely.
The case where $S$ and $S'$ are identical is discussed in
section~\ref{sec:sec_symm}.

\reduze\  is able to automatically detect sector relations or handle relations
supplied by the user.
Sector relations will be used to eliminate redundant sectors or integrals,
usually at the earliest possible stage.
As a particularly useful special case, a shift \eqref{eq:shift_trafo} might
map each propagator of an integral family to another propagator of the same
integral family.
This leads to a one--to--one mapping between integrals as well as sectors
of the integral family.
Such a relation can be entered for the full integral family via permutations
of propagators and allows for particularly efficient removal of redundancies.

As a generalization of the above described concepts,
\reduze\  also allows for crossings of
external momenta in addition to \eqref{eq:shift_trafo}.
If the involved crossing leaves the kinematic invariants
unchanged, the corresponding relations can be directly exploited for
relations between sectors as described before.
In the general case, \reduze\  also handles relations between sectors of
integral families defined with crossed kinematics.

\subsection{Sector symmetries}
\label{sec:sec_symm}

Special shifts of the loop momenta as in \eqref{eq:shift_trafo} which
transform a sector to itself are called \emph{sector symmetries}.
These shifts are also allowed to contain a permutation of the external
momenta as long as it does not change the kinematic invariants.
Sector symmetries may be used to express integrals in terms of other
integrals in the same sector and its subsectors.
These relations may provide information complementary to the IBP and LI
identities and can be used in the reduction to find a minimal number of
master integrals.
\reduze\ is capable of automatically determining sector symmetries
or handling user supplied rules, and offers to exploit them for reductions.

\section{Graph and matroid based algorithms for sectors}

\subsection{Physical sectors}
\label{sec:physical_secs}

A physical sector is a sector whose propagators correspond to edges in
a graph such that momentum is conserved. The construction of a graph from
a sector, i.e. from the momenta of a set of propagators with $l$-loop momenta,
can be done by choosing $l$ propagators which have independent momenta with
respect to the loop momenta and identifying them as edges in a graph with
both ends glued together in a single root vertex. External edges are also
attached to this root vertex with one of their ends.
Subsequently, for each of the remaining propagators, a vertex of the graph
(first the root vertex) is cleaved into two vertices, and a new edge is inserted
between these vertices, such that the edge's momentum (determined by momentum
conservation) exactly matches the propagator's momentum.

With this procedure \reduze\ automatically constructs graphs for sectors
where this is possible and thus identifies physical sectors.
The possibility of having graph representations for sectors gives
access to fast algorithms for identifying isomorphic graphs and
finding sector relations and sector symmetries.

\subsection{Sector relations}

If graphs constructed for two different physical sectors are isomorphic,
a shift of the form \eqref{eq:shift_trafo} between the two sets of loop momenta
can be derived by identifying the edges of the two graphs together
with their oriented momentum flow labeling.
\reduze\ offers the possibility to find relations between physical sectors
by this strategy, allowing also crossings of external legs for the graph
isomorphism but restricting to cases with $|\det{M}|=1$.
For the graph of each physical sector a standard form of its adjacencies,
a \emph{canonical label}, is computed with the algorithm
\cite{McKay1981}.
We distinguish different masses by replacing massive edges with a chain
of several edges, where the length of such a chain labels a mass uniquely.

\begin{figure}
\begin{center}
\includegraphics[width=0.25\textwidth]{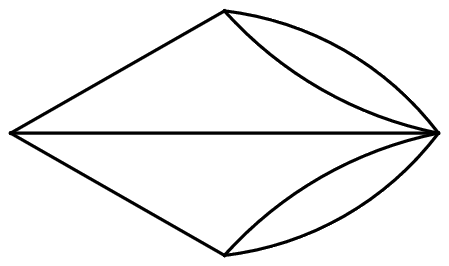}\hspace{20mm}
\includegraphics[width=0.25\textwidth]{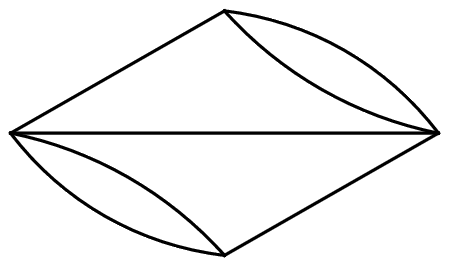}
\end{center}
\caption{\label{fig:twists}
Two non-isomorphic graphs which are related by a twist.
The matroids of these graphs are isomorphic.
}
\end{figure}

Graph isomorphism takes into account the ambiguities in labeling the
\emph{nodes} of a graph.
While isomorphic graphs can be described by the same propagators (possibly
with a crossing of external momenta), the inverse is not true.
Consider for instance the two vacuum graphs in figure~\ref{fig:twists}.
The two graphs are non-isomorphic but can be described by the same
propagators, i.e.\ by the same sector.
Here, a more appropriate object to consider is not the graph of the sector,
but the associated matroid.
Matroids are based on the notion of a set of linearly independent sets
and may be considered as generalizations of graphs.
For a graph an associated graph matroid (or cycle matroid) can be defined via
its edges. Definitions and fundamental properties are given in the review
\cite{Bogner:2010kv} and references therein.
Essential for us is the following chain of statements.
The relevant properties of a vacuum graph where all edges share a
common mass is encoded in the first Symanzik polynomial
($\mathcal{U}$ polynomial).
For brevity of the argument let us furthermore restrict to biconnected graphs.
The generalization to arbitrary vacuum components with different masses is
rather straightforward.
An immediate combinatorial approach to isomorphisms of the
Symanzik polynomials, which is not restricted to vacuum graphs, was presented in
\cite{Pak:2011xt}.
Here, we note that the first Symanzik polynomials
of two graphs are equal up to a permutation of edge variables exactly if
their matroids are isomorphic, see \cite{Bogner:2010kv} and references therein.
Two matroids of biconnected vacuum graphs are isomorphic exactly if the
two graphs are isomorphic up to a series of \emph{twists}, a statement known
as Whitney's theorem \cite{whitney}.
A twist operation starts by breaking a graph into two graphs such that
identification of \emph{separation pairs} of nodes in both graphs restore
the original graph.
As the second step of the twist, the separation pairs are identified
with flipped orientation.
In figure~\ref{fig:twists} the graphs are related by a twist around
the left-most and right-most nodes and thus have the same matroid.

These statements can be turned into an algorithm to detect shifts \eqref{eq:shift_trafo}
between vacuum sectors, which is sketched in the following and implemented in
\reduze.
We extend the graph isomorphism based shift detection by modifying the
generated graphs with twists such that their canonical labels are minimized.
A graph of a physical sector is decomposed into biconnected components with the
algorithm \cite{HopcroftT71}.
Possible separation pairs of biconnected components are identified via a
decomposition into \emph{triconnected components}.
We implemented the algorithm \cite{HopcroftT73, GutwengerMutzel2001}
for this purpose.
In order to generate at least one representative for each graph isomorphism
class, it is necessary to perform twists around separation pairs
as specified by \emph{virtual edges} as well as twists which correspond to
permutations of edges within polygon components of the decomposition.
While twisting we track the edges including their orientations to be able to
identify propagator momenta and different masses.
Graphs with external legs are handled by intermediately joining their external
nodes into one node and restricting to those twists, which keep the external
legs joined into one node with their original orientation.

Alternatively, \reduze\ also offers a procedure to find all sector relations,
which tries to identify sets of propagators in a straightforward approach
based on linear algebra and combinatorics.
While this procedure finds all shifts between arbitrary sectors,
including sectors not corresponding to a graph,
it is usually much less efficient than the graph based methods.

\subsection{Sector symmetries}
\label{sec:sym_rel}

Sector symmetries are shifts of the form \eqref{eq:shift_trafo} with
map a sector onto itself.
Additional permutations of the external momenta are permitted as long as the
kinematic invariants are unchanged.
Different such sector symmetries for physical sectors can be found by the
underlying symmetries of the graph as vertex permutations from the automorphism
group and permutations of multi-edges. The automorphism group of a graph
consists of all permutations of the vertices which leave the canonical label
of the graph unchanged. These transformations are calculated
in \reduze\ with the algorithm \cite{McKay1981} and are used to derive the
mapping of the edges between pairs of vertices and the associated shift of
the loop momenta. In the case where there is more than one edge between
two vertices also permutations of the edges with the same mass are considered.

Alternatively, a complete set of sector symmetries can be calculated
by \reduze\ using a combinatorial propagator matching approach.

\subsection{Matching of diagrams to sectors}
\label{sec:match_dias}

As discussed above, the assignment of momenta to propagators of Feynman
diagrams is ambiguous.
If diagrams are generated with a program like \qgraf\ \cite{qgraf}, typically
the assigned loop momenta have to be shifted in order to index the involved
loop integrals via integral families.
\reduze\ can compute these shifts for diagram files generated with \qgraf.
\reduze\ can also handle permutations of external momenta and find matchings of
diagrams to crossed sectors.

\section{Distributed reduction algorithm}

\subsection{Load balanced system solving}

\begin{figure}
\begin{center}
\begin{minipage}[b]{0.5\textwidth}
{\small
\fcolorbox{framecolor}{shadecolor}{
\parbox{52mm}{
\mbox{}\vspace*{-5mm}
\begin{align*}
 \textcolor{emphcolor}{\boldsymbol{I_5}} + c_{14} \boldsymbol{I_4} + c_{13} \boldsymbol{I_3} \qquad\quad\,\, \qquad\quad\,\,&= 0 \\
 \textcolor{emphcolor}{\boldsymbol{I_5}} + c_{24} \boldsymbol{I_4} \qquad\quad\,\,  + c_{22} \boldsymbol{I_2} \qquad\quad\,\,&= 0\\
 \textcolor{emphcolor}{\boldsymbol{I_5}} \qquad\quad\,\, + c_{33} \boldsymbol{I_3} + c_{32} \boldsymbol{I_2} \qquad\quad\,\,&= 0
\end{align*}
\mbox{}\vspace*{-7mm}
}}
\\[1mm]
\fcolorbox{framecolor}{shadecolor}{
\parbox{58mm}{
\mbox{}\vspace*{-5mm}
\begin{align*}
\qquad\quad\,\,  \textcolor{emphcolor}{\boldsymbol{I_3}} + c_{42} \boldsymbol{I_2} \qquad\quad\,\,  &= 0 \\
\qquad\quad\,\,\qquad\quad\,\, \textcolor{emphcolor}{\boldsymbol{I_3}}\qquad\quad\,\,  + c_{51} \boldsymbol{I_1} &= 0
\end{align*}
\mbox{}\vspace*{-7mm}
}}
\\[1mm]
\fcolorbox{framecolor}{shadecolor}{
\parbox{58mm}{
\mbox{}\vspace*{-5mm}
\begin{align*}
\qquad\quad\,\,\qquad\quad\,\,\qquad\quad\,\, \textcolor{emphcolor}{\boldsymbol{I_2}} + c_{61} \boldsymbol{I_1} &= 0
\end{align*}
\mbox{}\vspace*{-7mm}
}}
}\end{minipage}
\hspace*{5mm}
\includegraphics[width=0.35\textwidth]{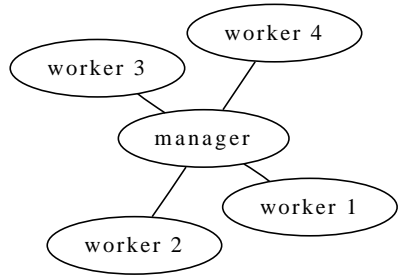}
\end{center}
\caption{\label{fig:blocks}
\emph{Left:} Three blocks of equations for loop integrals $I_i$
with coefficients $c_{ij}$ depending on kinematic invariants
and the space--time dimension.
Each block (shaded rectangle) contains equations
for the same leading integral (bold face $I_i$).
\emph{Right:} communication topology for MPI processes involved in a
distributed reduction.
}
\end{figure}

\reduze\ computes reductions for Feynman integrals by generating identities
for a range of seed integrals and reducing this linear system of equations.
The seed integrals are usually chosen for ranges in the propagator
exponent sums $r$ and $s$.
In a typical application, reductions for integrals from several sectors are
needed.
Moreover, a full reduction of a specific sector requires in general
also the reduction of subsector integrals.
\reduze\  proceeds bottom-up: the reduction of a sector is started only after
all subsector results are available.
Sectors which are no subsectors of each other can be reduced independently such
that these tasks are easily parallelized.
This kind of parallelization is available in the first version of \reduze\ 
via a shell script, which launches programs for different sectors.
In version\ 2 of \reduze, two levels of parallelization are implemented via the
message--passing--interface (MPI) standard.
The reduction of a sector becomes a \emph{job} and
several independent such reduction jobs can be executed in parallel.
The job system is described in more detail in the next section.
On top of this first level of parallelization, this version of \reduze\  implements
a distributed reduction algorithm for the reduction of a single sector.
Since in this case the parallelization is less obvious we give some details about
our implementation in the following subsection.

In \reduze, a total ordering is defined for indexed integrals of the type
\eqref{eq:int_reduze_format}. The ordering defines integrals of a sector to be more
complicated than integrals of its subsectors.
In the following, terms like ``leading'' or ``lower'' integral refer
to this ordering.
To choose specific integrals as master
integrals the user may adjust the ordering, possibly at a later stage.
In order to reduce integrals of a given sector, IBP and LI identities are
generated for a specified range of seed integrals.
This results in a sparse homogeneous linear system of equations for the indexed
integrals where the coefficients are rational functions of the kinematic
invariants and the space--time dimension.
The equations are sorted into blocks
of equations with the same leading integral,
see left panel of figure \ref{fig:blocks}.

The blocks are reduced bottom up, starting from the block with the lowest
leading integral.
For each block, integrals are reduced, i.e. replaced by linear combinations of
lower integrals, according to the results from lower blocks and subsectors
(``back substitution'') if possible.
If a block contains several equations, one is kept and used to eliminate
the leading integral from all other equations in the block (``forward
elimination''), which are subsequently solved for the new leading integral.
The coefficients of the integrals are normalized such that zeros are detected
and numerator and denominator are coprime.
This requires multivariate polynomial greatest--common--divisor (GCD)
computations which typically present the most time consuming part of the
full calculation.
The result of this reduction of a block is one equation for the block and
possibly further equations with lower leading integrals to be inserted
into lower blocks.
The next block to be selected is the lowest block which contains more than
one equation or involves integrals which can be reduced.

\reduze\  offers both, a purely serial reduction for one sector as well as
a distributed execution.
In the serial version, the above steps are performed in deterministic order
on one core.
The distributed version employs a star topology of MPI processes with one
manager and one or more workers, see right panel of figure~\ref{fig:blocks}.
The workers perform the actual reduction steps for a block, while
the manager keeps track of the complete system of equations and balances the
work between the workers.
More specifically, an idle worker contacts the manager to request work.
The manager looks up the next block to process and sends its equations together
with equations needed for back substitutions to the worker.
The worker reduces the block and sends the result to the manager.

Our motivation for this distributed algorithm is as follows.
Experiments show that in typical applications the time needed for the reduction
of one block can easily differ by more than 6 orders of magnitude.
Moreover, the exact order of the individual reduction steps significantly
determines the execution time for the full reduction and a bottom-up order
typically shows the best performance.
Both issues are directly addressed by the dynamical load balancing mechanism
presented above, at least for a not--too--large number of worker processes.
How well this works in practice is quantified in the following subsection.

\reduze\ allows to choose between \ginac\ \cite{Bauer:2000cp}
and \fermat\ \cite{fermat} for the GCD computation
needed to normalize coefficients.
During reduction, the equations are stored either in RAM or optionally
in a transactional database as implemented by the open source Berkeley DB
\cite{berkeley}.
With transactions turned on, an aborted reduction of a single sector may be
resumed at a later time; recovering completed reductions for sectors is
available in any case.
\reduze\ supports different ways to split a calculation into several runs,
this is described in the tutorial provided with the \reduze\ distribution.
Reductions for crossed integrals are automatically be obtained via reduction
results from its uncrossed counterpart in order to save computation time
and disk space.

\subsection{Performance results for single sectors}
\label{sec:perf_single}

\begin{figure}
\centerline{\includegraphics[width=0.5\textwidth]{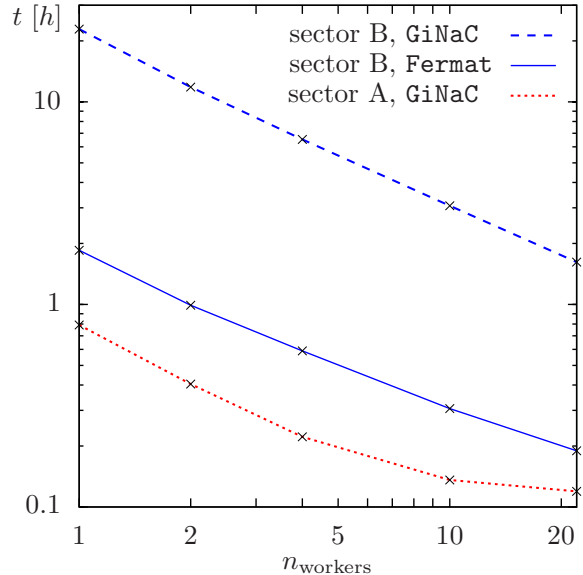}}
\caption{\label{fig:perf_single}
Execution time for the reduction of single sectors in dependence
of the number of worker processes employed for the reduction.
}
\end{figure}

Performance results for the reduction of sectors for two--loop contributions
to heavy--quark pair production are shown in figure~\ref{fig:perf_single}.
These sectors have $t=4$ or $5$, respectively,
and reductions were computed for integrals with $r=t\ldots 7$ and
$s=0\ldots3$ or 4, respectively.
We used a computer with 48 CPU cores operating at $2.1$~GHz and started
\reduze\ with $n_{\text{workers}}+2$ MPI processes (one job center process
and one manager process should be overbooked for a small number of available
cores).

In general, we observe that the scaling with the number of processes is
problem specific and depends on the configuration of \reduze, such as the
chosen computer algebra system.
The upper two curves in the figure show an example with a good scaling for up
to 22 workers.
Indeed, we find examples where the scaling is good up to 48 workers.
We think this good scaling behavior is noteworthy, given the fact that it
describes the distributed computation of a not too loosely coupled system.
As expected, we observe that beyond some number of worker
processes the run time decreases less and less with additional workers
and finally increases for even larger number of workers.
Contributions to this behavior is expected from serial parts
of the code, communication overhead, but potentially also from
a ``less ideal'' order of evaluation when solving the system of equations
with a larger number of workers. 
The lowest curve was obtained for an example of a reduction of a comparably
simple system of equations, where the onset of such a behavior is clearly
visible.
It is also not difficult to find examples with worse scaling,
where a minimal runtime is obtained for only a few worker processes.

Using \fermat\ instead of \ginac\ for the GCD computations can easily result
in a speed-up by an order of magnitude, see the two blue curves for
sector~B in the figure.
For the displayed \fermat\ benchmarks, the system to be reduced was kept in RAM,
while for the \ginac\ benchmarks a database was used.
Our tests indicate that for examples of this type, the performance penalty for
using a database is considerably less than the differences due to the two
algebra systems considered here.
Nevertheless, also this option is relevant for optimal performance,
especially for a larger number of workers.
Typically, if the coefficients in the equations become more involved, the
reduction benefits considerably from a larger amount of workers compared
to cases with compact coefficients.
The performance in a real application example, involving the reduction of
several sectors, is discussed at the end of the next section.

\section{Job system}

\subsection{Load balanced execution of jobs}

\begin{figure}
\centerline{\includegraphics[width=0.8\textwidth]{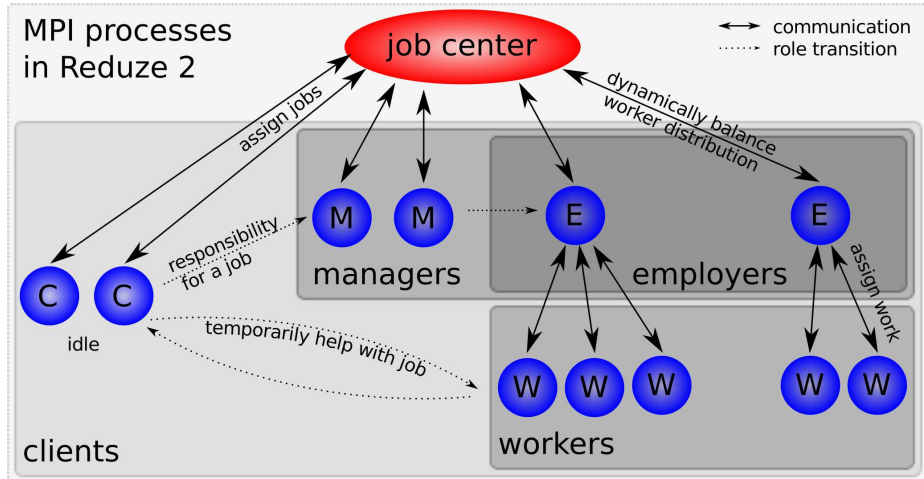}}
\caption{\label{fig:jobcenter}
Dynamical load balancing in \reduze}
\end{figure}

In \reduze, a \emph{job} represents a sequence of computations which
can be performed once its dependencies, specified via the presence of input
files, are fulfilled.
Most jobs in \reduze\ are \emph{serial jobs}: they are executed on a single
core but possibly in parallel to other jobs.
Parallel execution of different jobs represents the top layer parallelization
mechanism of \reduze\ which is automatically available for any job type added
to \reduze.
Reduction of identities is an example for a \emph{distributed job}.
Such a job can be executed by several processes in parallel: one process,
the \emph{manager}, is responsible for the full execution of the job,
other processes, the \emph{workers}, help for some time with the execution.
In order to employ this second layer parallelization, a dedicated distributed
algorithm needs to be implemented for the job.
For each run of \reduze, the user specifies a sequence of such jobs, which
are inserted into a job queue. A job may generate additional auxiliary jobs
automatically. The job queue is responsible for resolving the dependencies
between the jobs and determining the next job to be executed.

If \reduze\ is started with several MPI processes, one process will act as a
\emph{job center} and dynamically balance work between the other processes,
the \emph{clients} of the job center, see figure~\ref{fig:jobcenter}.
The job center schedules jobs using the job queue and assigns them to clients.
An idle client contacts the job center and requests work.
The job center responds by assigning a job to the client, either to be executed
as a manager or as a worker.
The client changes its role accordingly.
As a worker, it contacts the responsible manager of the job and helps with the job
execution.
As a manager, it can either execute the job by itself or
register as an \emph{employer} at the job center to request workers.

In order to optimize the efficiency of the parallelization,
the job center periodically collects performance measurements from the employers
and estimates an optimal distribution of workers based on it.
According to this estimate, workers will effectively be reassigned to other
employers by requesting release of workers from ``inefficient'' employers and
assigning idle customer to ``efficient'' employers.
The basic idea is to avoid low efficiencies due to overloaded managers by
assigning workers preferably to managers which idle a high percentage of their
CPU time.

\subsection{Performance results for multiple sectors}

\begin{figure}
\begin{center}
\begin{minipage}[b]{0.21\textwidth}
 \includegraphics[width=\textwidth]{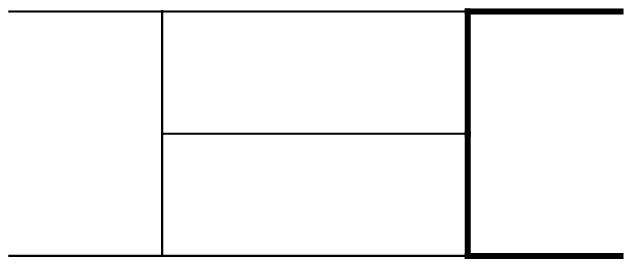}\\[2ex]
 \includegraphics[width=\textwidth]{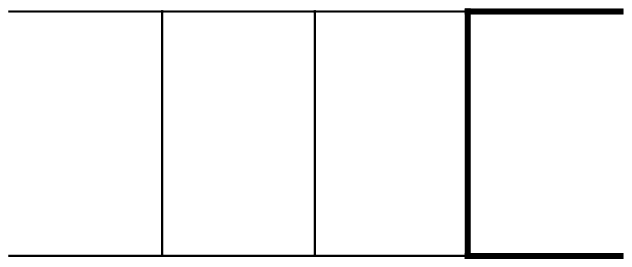}\\[2ex]
 \includegraphics[width=\textwidth]{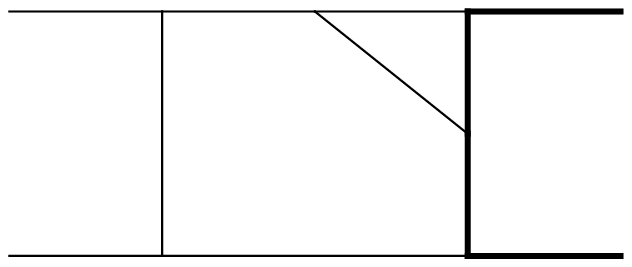}\\[2ex]
 \includegraphics[width=\textwidth]{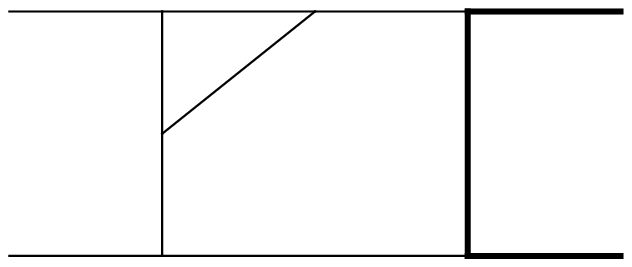}
\vspace{2ex}
\end{minipage}
\hspace{10ex}
\includegraphics[width=0.5\textwidth]{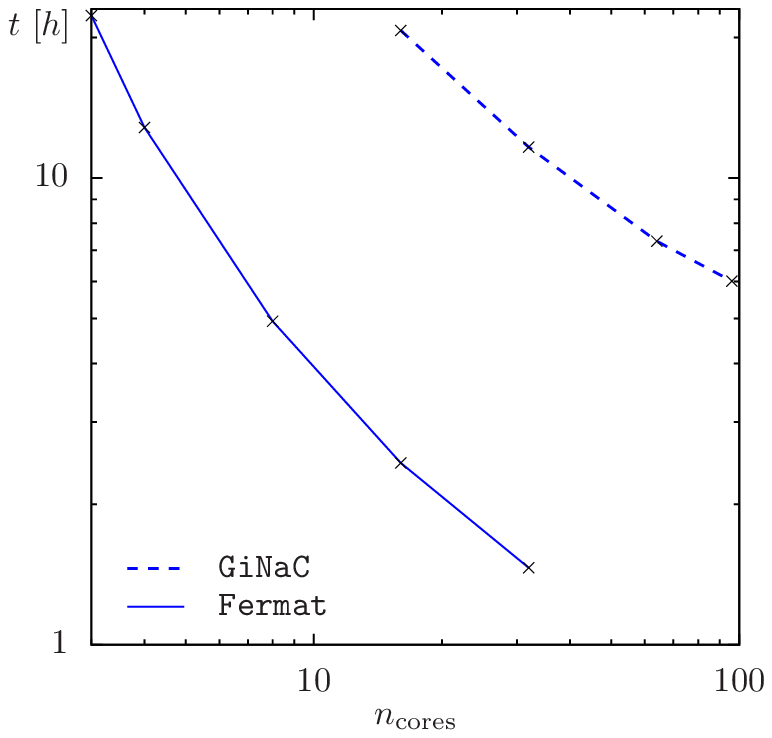}
\end{center}
\caption{\label{fig:perf_multi}
Runtime decrease for the reduction of multiple sectors using a large number
of processor cores.
The reductions cover the sectors of the depicted graphs and all of their
subsectors.
In the graphs, thick lines represent massive propagators.
}
\end{figure}

Performance results for the reduction of a selection of sectors are shown in
figure~\ref{fig:perf_multi}.
These sectors are encountered in the calculation of two-loop corrections to
heavy-quark pair production.
Reductions are calculated for integrals of the depicted sectors and all of
their subsectors, where $r=t\ldots 7$ and $s=0\ldots 4$.
The benchmarks for the solid curve were performed on a computer with 48 CPU
cores operating at $2.1$~GHz using \fermat\ and keeping all equations in RAM.
The measurements for the dashed curve were obtained using the ``Schr\"odinger''
cluster of the University of Zurich with $2.8$~GHz cores; these runs were
configured to use \ginac\ and a database.

The figure shows for this realistic application example that the calculation
benefits considerably from up to 96 cores, if available.
Let us stress again, that the scaling behavior is problem specific and
may be worse for other types of applications.
In the present example,
the run--time decrease due to additional cores is quite close to an optimal
$1/n_{\text{cores}}$ behavior for smaller numbers of cores.

\section{Other features}

\subsection{Differential equations for master integrals}
One method to solve master integrals consists
of deriving differential equations by taking derivatives in the kinematic
invariants, replacing new integrals with the reduction results and solving
the differential equations by integration.
In particular for sectors with several independent integrals,
a change of basis for the master integrals may be required.
\reduze\ offers the possibility to derive differential equations for Feynman
integrals and reduce these equations for some user choice of master integrals.

\subsection{Interference terms}
Starting from diagrams generated by \qgraf, \reduze\ can compute scalar
interferences of (bare) QED or QCD amplitudes in dimensional regularization.
This includes insertion of user--defined Feynman rules, contraction of Lorentz
vector indices, performing Dirac traces, and evaluating color structures.
In case of an interference of a tree-level diagram with a diagram that
could be matched to a sector of an integral family (section \ref{sec:match_dias})
also the occurring integrals, which belong to the matched sector and its subsectors,
are indexed by the corresponding integral family.
In a further step, these integrals can be reduced to master integrals and
substituted correspondingly in the interference terms.
Each computation of an interference of two diagrams is treated as a job, and
when MPI is used, these jobs can be performed in parallel.
Specific examples are distributed with the \reduze\ package.

\section{Usage}

The package \reduze\ can be downloaded from the web page
 \url{http://projects.hepforge.org/reduze}\,.
The distribution contains a tutorial with detailed description of
installation and usage as well as several example files.

\acknowledgments{
We thank York Schr\"oder for many useful discussions and comments on the
draft.
We are grateful to the CEDAR collaboration for hosting our project web page
on HepForge~\cite{Buckley:2006np}.
This work was supported by the Deutsche Forschungsgemeinschaft (DFG SCHR 993/2-1)
and the Schweizer Nationalfonds (Grant 200020\_124773/1).}

\end{document}